\documentstyle[aps,prb,multicol,graphics,epsf]{revtex}
\begin{document}
\draft

\title{Mott-Hubbard insulators for systems with orbital degeneracy}
\author{Olle Gunnarsson$^{(a)}$, Erik Koch$^{(b)}$,
        and Richard M. Martin$^{(b)}$}
\address{${}^{(a)}$Max-Planck-Institut f\"ur Festk\"orperforschung,
         D-70506 Stuttgart, Germany}
\address{${}^{(b)}$Department of Physics, University of Illinois at
         Urbana-Champaign, Urbana, Illinois 61801}

\date{\today}

\maketitle

\begin{abstract}
We study how the electron hopping reduces the Mott-Hubbard band gap
in the limit of a large Coulomb interaction $U$ and as a function of the
orbital degeneracy $N$. The results support the conclusion that the 
hopping contribution grows as roughly $\sqrt{N}W$, 
where $W$ is the one-particle band width, but in certain
models a crossover to a $\sim NW$ behavior is found for sufficiently
large $N$. 
\end{abstract}
\pacs{71.10.Fd, 71.30.+h}

\begin{multicols}{2}
\section{Introduction}
The Mott-Hubbard metal-insulator transition has attracted much interest     
since it was proposed by Mott,\cite{Mott} being an example of how 
strong correlation drastically changes the properties of the system. 
For simplicity, the Mott-Hubbard transition is usually studied for
models without orbital degeneracy,\cite{Georges} e.g., the Hubbard
model.\cite{Hubbard} For most real systems, however, the orbitals 
primarily involved in the transition are degenerate. We recently 
argued that in a Hubbard model with an orbital degeneracy $N$, the ratio
of $U/W$ where the Mott-Hubbard transition takes place is increased
by roughly a factor $\sqrt{N}$,\cite{deg} where $U$ is the on-site 
Coulomb interaction and $W$ is the one-electron band width.

This conclusion was based on a simple and suggestive but nonrigorous
 argument in the large $U$-limit, which said that the band gap $E_g$  
behaves as\cite{deg}
\begin{equation}\label{eq:1}
E_g\sim U-\sqrt{N}W.
\end{equation}
This argument was supported by exact diagonalization calculations 
for small clusters with $N=1$, 2 and 3. 
The argument    was then extrapolated to intermediate values 
of $U$, and the  extrapolation was supported by quantum lattice 
Monte Carlo calculations for a model of A$_3$C$_{60}$ (A=K, Rb) 
with the $N=3$.
Similar results have recently also been obtained in Monte Carlo calculations 
by Han and Cox,\cite{Han} while Lu\cite{Lu} found a $(N+1)W$ behavior 
using the Gutzwiller Ansatz and Gutzwiller approximation.

Here we want to investigate the behavior in the large $U$ limit further.
We have in particular studied a model with hopping only between orbitals
with the same  orbital quantum number, referred to as an intraband 
hopping model.
This model shows an interesting cross-over as the degeneracy $N$ becomes 
larger than the number of nearest neighbor $K$. Thus the hopping
contribution tends to be $\sim\sqrt{N}W$ for $N<<K$ but $\sim NW$ for $N>>K$.
When hopping between orbitals with different quantum numbers is 
included, interband hopping models,
the results become more model dependent. The results obtained
here support a behavior $\sim \sqrt{N}W$ or a somewhat more rapid growth
with $N$.

In Sec. II we present some general arguments. In Sec. III we  
use exact diagonalization to study small clusters with intraband 
hopping and in Sec. 
IV this is extended to include interband hopping. The results are
discussed in Sec. V.

\section{General considerations}
We  consider a system with $M$ sites and the orbital degeneracy $N$ at 
half-filling, i.e., with $NM$ electrons.
The band gap is then given by
\begin{equation}\label{eq:3}
E_g=E(NM+1)+E(NM-1)-2E(NM),
\end{equation}
where $E(L)$ is the ground-state energy for   $L$ electrons. 
In the large $U$-limit, the $NL$-electron ground-state has exactly 
$N$ electrons 
per site. The hopping of an electron to a neighboring site would
cost the Coulomb energy $U$ and is strongly suppressed in this limit.
Therefore 
\begin{equation}\label{eq:4}
  E(NM)={1\over 2}N(N-1)MU + {\cal O}\left({t^2\over U}\right),
\end{equation}
where $t$ is a hopping integral.
For the states with an extra electron or hole the situation is more 
complicated, since the extra electron or hole can hop without any extra cost 
in Coulomb energy. We therefore focus on these states. We first form 
an antiferromagnetic state $|{\rm anti}\rangle$ with $N$ electrons of 
a given spin on each site and with each site having as many neighboring 
sites of the opposite spin as possible. To be specific, the
cental site 1 has spin up. We then form a state with $NM+1$ electrons
\begin{equation}\label{eq:5}
|v_1\rangle=\psi^{\dagger}_{11\downarrow}|{\rm anti}\rangle,
\end{equation}
where $\psi^{\dagger}_{im\sigma}$ creates an electron on site $i$ in
the orbital $m$ with spin $\sigma$.
Thus site 1 in the state $|v_1\rangle$ has an additional electron.
A spin up electron can hop from this site to a neighboring spin down site,    
forming a state 
\begin{equation}\label{eq:6}
|v_2\rangle={1\over \sqrt{NK}}\psi^{\dagger}_{11\downarrow}
\sum_{im}\psi^{\dagger}_{im\uparrow}
\psi_{1m\uparrow}|{\rm anti}\rangle  ,
\end{equation}
where the sum $i$ is over the $L$ ($\le K$) neighboring sites 
with opposite spin to site 1. 
The spin down electron on site 1   can also hop to neighboring sites
with spin up, but we neglect this for the moment. The matrix element
between the states in Eqs. (\ref{eq:5}-\ref{eq:6}) can then be calculated.
\begin{equation}\label{eq:7}
\langle v_2|H|v_1\rangle=\sqrt{NL}t,
\end{equation}
where we have considered the case when there is a hopping matrix element 
only between orbitals of the same $m$ quantum number on the neighboring
sites. The argument can also be repeated for more general hopping.\cite{deg}
The important aspect of the result (\ref{eq:7}) is that it has a factor
$\sqrt{N}$ relative to the one-particle case.\cite{deg,Ce}
 This is a consequence of the
fact that any of the $N$ spin up electrons can hop to the neighboring
site, as illustrated in Fig. \ref{fig1}.
For a bipartite system with $N=1$, the ferromagnetic arrangement 
in Fig. \ref{fig1}a is favorable according to Nagaoka's theorem.\cite{Nagaoka}
For $N>1$ the situation in Fig. \ref{fig1}b is in general more favorable,
since the hopping can then take place in $N$ channels.

\noindent
\begin{minipage}{3.375in}
\begin{figure}
  \centerline{\epsfxsize=2in \epsffile{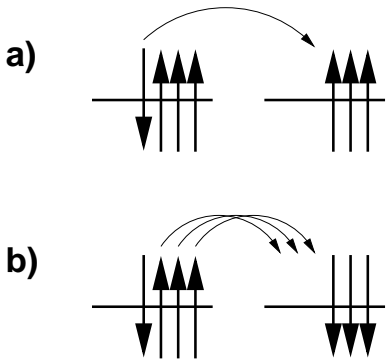}}
  \hspace{2ex}
  \caption[]{\label{fig1}
    Illustration how an extra electron can hop against a half-filled background.
    a) ferromagnetically aligned neighbor: There is only one hopping-channel.
    b) antiferromagnetically aligned neighbor: There are $N$ different ways
       for hopping.}
\end{figure}
\vspace{1ex}
\end{minipage}

For one single electron in the system, the hopping energy is of the order 
of $W/2$, where $W$ is the one-electron band width. The extra factor $\sqrt{N}$
in Eq.~(\ref{eq:7}) suggests that the hopping energy is correspondingly
larger for the $NM+1$ electron case and that the total energy is
\begin{equation}\label{eq:8}
E(NM+1)\approx E(NM)+NU+\sqrt{N}W/2.                 
\end{equation}
Analogous arguments then suggest   
a similar result for $E(NM-1)$ and we obtain 
\begin{equation}\label{eq:9}
E_g=U-\sqrt{N}W.                                                
\end{equation}

As pointed out in Ref. \onlinecite{deg}, these arguments are not rigorous.
As the extra occupancy moves through the system it reduces the spins
along its path. There is therefore no one to one correspondence to the 
one-particle case. We therefore need to consider the arguments in more
detail.
 
To be specific, we consider a degenerate multiband Hubbard model
\begin{eqnarray}\label{eq:10}
  H &&= \sum_{<ij>m\sigma} t_{im,jm^{'}}\,\psi^{\dagger}_{im\sigma}
\psi_ {jm'\sigma}  \nonumber  \\
   && + U\sum_{i\;(m \sigma) < (m'\sigma')}n_{im\sigma}n_{im'\sigma'},
\end{eqnarray}
with hopping only between orbitals on nearest neighbor sites.
In part of the discussion we furthermore consider the intraband 
hopping model, where $t_{im,jm^{'}}
\equiv t\delta_{mm^{'}}$, i.e., only hopping between orbitals with the
same quantum number $m$.                               

To calculate the energies of the states with an extra hole or electron,
we use the Lanczos method.\cite{Lanczos}
In this method a set of basis states $|v_n\rangle$ are generated 
according to the prescription
\begin{eqnarray}\label{eq:11}
  &&b_{n+1}|v_{n+1}\rangle = H|v_n\rangle-a_n|v_n\rangle-b_n|v_{n-1}\rangle\\
  &&a_n = \langle v_n |H|v_n\rangle  \nonumber
\end{eqnarray}
where $b_{n+1}$ is determined by the requirement that $|v_{n+1}\rangle$
is normalized. In this way the Hamiltonian is transformed into a tridiagonal
form, where the diagonal elements are given by $a_n$ and the nondiagonal
elements by $b_n$ ($H_{nn}=a_{n}$ and $H_{n,n+1}=b_{n+1}$). The starting state 
$|v_1\rangle$ is defined in Eq.~(\ref{eq:5}) and $|v_{0}\rangle\equiv 0$.

This approach gives the energy of the lowest eigenstate which is  
nonorthogonal to the state $|v_1\rangle$.  
If our choice of $|v_1\rangle$ is orthogonal to the ground-state of the
$(NM\pm 1)-$system, the energies of these ground-states are
overestimated. We then obtain an upper limit to $E_g$,
since we know the energy of the half-filled state in the large $U$-limit.
This remains 
true if the iterative steps implied by Eq.~(\ref{eq:11}) are interrupted
before convergence, since the lowest eigenvalue obtained from the tridiagonal
matrix is decreased as more (nonzero) elements are added to the matrix.
We furthermore observe that if $|v_1\rangle$ is chosen poorly 
in the sense that it is
very different from the ground-state, the ground-state energy is still obtained
after sufficiently many steps, as long as $|v_1\rangle$ is not orthogonal to
the ground-state.

In the following we often use a classical N$\acute{\rm e}$el 
state to construct the 
states $|{\rm anti}\rangle$ and $|v_1\rangle$. These states are 
convienent to handle, but not necessarily the best starting point, 
due to the tendency of the system to form a singlet (or a doublet with
an extra electron or hole for $N>1$). Due to the arguments above 
this will at worst mean that we underestimate the hopping reduction
of the gap. 

In the case of a half-filled system with an additional hole or electron,
the Coulomb energy is minimized if there are exactly $N$ electrons per
site, except for one site with the extra electron or hole. Since we consider
the large-$U$ limit we allow only such states in the following. 
The Coulomb energy is then a constant which
we choose as energy-zero. We thus arrive at a $t$-$J$ model with $J=0$. Let
$T_{\rm lo}(NM\pm1)$ and $T_{\rm hi}(NM\pm1)$ denote the lowest and highest 
eigenenergy of the Hamiltonian.
In the large $U$-limit
\begin{equation}\label{eq:12}
  T_{\rm lo}(NM\pm 1)=-T_{\rm hi}(NM\mp 1).
\end{equation}
It is therefore sufficient to study, e.g., the $(NM+1)$-electron case, 
which gives
\begin{equation}\label{eq:13}
  E_g=U+T_{\rm lo}(NM+1)-T_{\rm hi}(NM+1).
\end{equation}

\section{Intraband hopping}
We have  studied a few small clusters of atoms, where the
system is small enough to allow an exact diagonalization.
We first consider intraband hopping models.
In these models the one-particle band width of each channel 
$m$ is the same and independent of $N$.

\subsection{Diatomic molecule}
We first study the case of just two sites, orbital degeneracy $N$ and
$2N+1$ electrons. We form one state $|a\rangle$, with $N+1$ electrons
on site 1. This state is a linear combination of states, where the
first state has $N$ spin up electron and one spin down electron 
with, say, $m=1$ on site 1 and $N$ spin down electrons on site 2. 
To generate further states in $|a\rangle$,
we let one of the spin up electrons hop to site 2 and one spin down  
electron on site 2 hop to site 1. This gives $N(N-1)$ states which are 
added to state $|a\rangle$. In this     way we form all possible states  
with $N+1$ electrons on site 1 and include them in $|a\rangle$.
In this process we also generate all possible states with $N+1$ electrons
on site 2, which are included in a state $|b\rangle$. For symmetry reasons,
state $|a\rangle$ and $|b\rangle$ include equally many terms $L$ and therefore
have the same normalization factor $1/\sqrt{L}$. 

 In calculating the matrix
element $\langle b|H|a\rangle$, we notice that in each $(m,\sigma)$ channel 
except $(m=1,\downarrow)$, there is only one electron.
Each one of the $L$ states in $|a\rangle$ then connects to exactly 
$N$ states in $|b\rangle$,
since $N$ of the electrons (spin up or down) on site 1 can hop to site 2,
 and only the electron in the orbital $(m=1,\downarrow)$ cannot hop.
There are then $NL$ contributions $t$ to
the matrix element. After taking the normalization factor into account,
we then find that the matrix element is $Nt$. The corresponding result
for the gap in the large $U$ limit is then
\begin{equation}\label{eq:d1}
E_g=U-2Nt=U-NW.
\end{equation}
The reduction of the band gap due to hopping is thus proportional to $N$
and not $\sqrt{N}$ as the arguments in the previous section suggested. 
If we start the Lanczos procedure from the antiferromagnetic state 
discussed before (Eq.~(\ref{eq:5})), $b_2=\sqrt{N}t$. 
For $n\sim N$, however, $b_n\sim Nt$.
The ground-state of the $(2N+1)$-electron system is a doublet. It is then 
not surprising that a large number of Lanczos steps is needed to generate
this state from antiferromagnetic state $|v_1\rangle$ in Eq.~(\ref{eq:5}).

From the results above it is immediately clear that 
for sufficiently large values of $N$, the reduction of the band gap due
to hopping must at least be of the order $Nt$ even for a larger system. 
This can bee seen by simply, arbitrarily, picking two neighboring atoms
in the system of interest and then constructing a state like above. 
This already gives a reduction of the band gap $\sim Nt\sim NW$, and any 
improvement of this simple construction can only make the reduction 
larger.

\subsection {``Bethe'' lattice}  
To obtain a system where an atom has several neighbors,
we have studied a simple version of a finite Bethe lattice, where 
atom 1 connects to $K$ neighboring atoms, while these atoms only
connect back to the first atom and has no other neighbors.
We have solved this problem using exact diagonalization for a number of 
values of $N$ and $K$ and find that the results are described by
\begin{equation}\label{eq:b1}
Eg=\cases {U-\sqrt{N+(N^2-1)/K}W & if $K> 1$;\cr
           U-NW  & if $K=1$.\cr}
\end{equation}
This result shows an interesting cross over behavior. For $N<<K$,
the reduction of the band gap is proportional to $\sqrt{N}W$, as suggested
by the argument in Sec. II. On the other hand, for $N>>K$ the reduction
is proportional to $NW$, as suggested by the previous section (IIIA). 
The limit $N<K$ should apply to most cases of practical interest, 
but the result for $N>>K$ is also interesting.

\noindent
\begin{minipage}{3.375in}
\begin{figure}
\rotatebox{270}{  \centerline{\epsfxsize=1.2in \epsffile{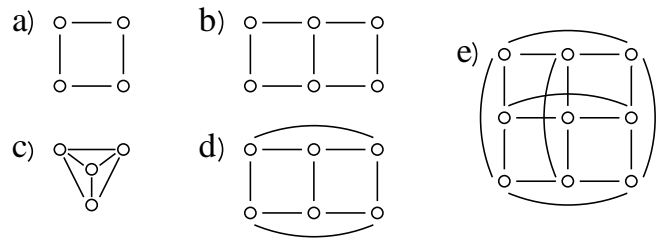}}}
  \hspace{2ex}
  \caption[]{\label{fig2} Schematic picture of small clusters studied by
exact diagonalization. Each circle indicates a site and the lines 
show the hopping between these sites.  }
\end{figure}
\vspace{1ex}
\end{minipage}

\subsection{Other small clusters}
The results for the Bethe cluster are instructive, but because of 
the simplicity of the model it is not clear  to what extent they apply to more 
realistic models. We have therefore studied a few models which 
are small enough to allow exact diagonalization. 
In Fig. \ref{fig2} we show these models schematically.
We define a quantity
\begin{equation}\label{eq:s1}
C(N)={ {\rm lim} \atop U\to \infty}{U-E_g(N)\over U-E_g(N=1)}{1\over \sqrt{N}},
\end{equation}
where $E_g(N)$ is the band gap for the orbital degeneracy $N$.
$C(N)\equiv 1$ implies that hopping reduction of the band gap is        
proportional to $\sqrt{N}$. The results are shown in Fig.~\ref{fig3}.
The figure supports the results found in Sec. III for the ``Bethe'' 
lattice, in the sense that $C(N)$ is closer to  a constant the larger $K$
is, but that $C(N)$ grows when $N\sim K$.

\noindent
\begin{minipage}{3.375in}
\begin{figure}
  \centerline{\epsfxsize=3.3in \epsffile{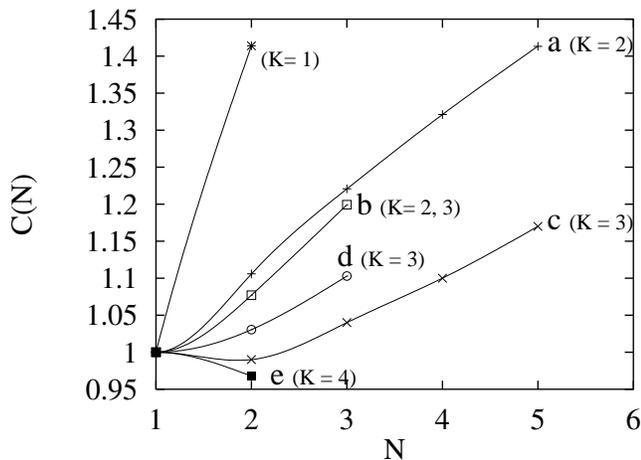}}
  \hspace{2ex}
  \caption[]{\label{fig3} The ratio $C(N)$ between $U-E_g$ for a 
given value of the orbital degeneracy $N$ and the same quantity for $N=1$
divided by $\sqrt{N}$. Thus if the reduction of the band gap were
proportional to $\sqrt{N}$, $C(N)$ would be identically unity.
The figure illustrates how this tends to be true when the number $K$ of
nearest neighbors is much larger  than $N$. The labels refer to
Fig. \ref{fig2}. We have also included the results for a diatomic
molecule ($K=1$). }
\end{figure}
\vspace{1ex}
\end{minipage}

\subsection{Tetrahedron}
In this section we study the tetrahedron ( c) in Fig. \ref{fig2}) 
in somewhat more detail.
This system has four sites
with equal hopping matrix elements $t<0$ between all the sites.
In the one-particle case, there is one nondegenerate eigenvalue $3t$
and a three-fold degenerate eigenvalue -$t$. The band width is then
\begin{equation}\label{eq:15}
W=4t.
\end{equation}
We notice that the band width is reduced by the frustration, since it 
is not possible to find a fully anti-bonding state.

 It is  
instructive to obtain this result from the Lanczos method. The nonzero 
coefficients $a_n$ and $b_n$ are shown in Table~\ref{tableI}.
The process stops already after two step, since 
\begin{equation}\label{eq:16}
H|v_2\rangle-a_2|v_2\rangle-b_2|v_1\rangle\equiv 0.
\end{equation}
We note that in the present case an electron on a given site can hop to four
other sites. The effect of one such hop is cancelled by the term $b_n|v_{n-1}
\rangle$. The remaining hops may contribute to $b_{n+1}$ or to $a_n$. In the
present case the contribution goes to $b_{n+1}$ in the first step but to
$a_n$ in the second step. For the many-body case the coefficients        
$a_n$ are typically smaller and a larger part of the hopping contribution 
instead goes to $b_{n+1}$.

For $N=1$, 2 and 3 the lowest possible $(NM\pm 1)$-energy is reached after 
just a few (4-8) Lanczos steps, illustrating that the choice 
in Eq.~(\ref{eq:5}) of $|v_1\rangle$ leads to the ground-states.      
The situation is different for $N=\infty$. We have calculated the first 
40 coefficients $b_{n+1}$, of which the first few are shown in
Table \ref{tableI}. At least up to $n=40$ $b_{n+1}/\sqrt{N}$ keeps
increasing with $n$. This suggests that the reduction of the band gap due 
to hopping grows faster that $\sqrt{N}$, in agreement with the arguments
in Sec. III.A-B that a contribution $\sim N$ may be expected for $N>>K$.

\noindent
\begin{minipage}{3.375in}
\begin{table}[h]
  \caption[]{The coefficients $a_n$ and $b_n$ for a tetrahedron. Both the
     one-particle and $(NM-1)$ many-particle problems are considered, where 
     in the latter case the orbital degeneracies $N=$ 1, 2, 3 and $\infty$ 
     are considered. The coefficients are given in units of $t\sqrt{N}$, 
     where $t$ is the hopping matrix element.}
  \hspace{2ex}
  \begin{tabular}{ccccccccccc}
    $n$  & \multicolumn{2}{c}{One-particle} &\multicolumn{8}{c}{Many-particle}\\
    & & & \multicolumn{2}{c}{$N=1$} & \multicolumn{2}{c}{$N=2$} &
      \multicolumn{2}{c}{$N=3$} & \multicolumn{2}{c}{$N=\infty$}   \\
    & $a_n$ & $b_{n+1}$ & $a_n$ & $b_{n+1}$ &$a_n$ & $b_{n+1}$ 
      &$a_n$ & $b_{n+1}$ &$a_n$ & $b_{n+1}$ \\ 
    \tableline
    1 & 0.00 & 1.73 & 0.00 & 1.73 & 0.00 & 1.58 & 0.00 & 1.53 & 0.00 & 1.41 \\
    2 & 2.00 & 0.00 & 0.67 & 1.25 & 0.57 & 1.30 & 0.49 & 1.32 & 0.00 & 1.41 \\
    3 &      &      & 0.19 & 1.51 & 0.44 & 1.48 & 0.45 & 1.49 & 0.00 & 1.58 \\
    4 &      &      & 0.25 & 1.07 & 0.31 & 1.21 & 0.24 & 1.34 & 0.00 & 1.76 \\
    5 &      &      &-0.11 & 0.00 &-0.17 & 0.98 &-0.00 & 1.37 & 0.00 & 1.94 \\
    6 &      &      &      &      & 1.15 & 0.97 & 0.55 & 1.21 & 0.00 & 1.96 \\
    7 &      &      &      &      &-0.17 & 0.00 &-0.20 & 0.92 & 0.00 & 2.24 \\
    8 &      &      &      &      &      &      & 0.94 & 0.91 & 0.00 & 2.17 \\
    9 &      &      &      &      &      &      &-0.16 & 0.00 & 0.00 & 2.46 \\
  \end{tabular}
  \label{tableI}
\end{table}
\end{minipage}

To discuss the large $N$-limit, we first discuss the general situation
when each site is surrounded by $L(\le K$) other sites with the opposite
spin. To leading order in $N$ there are then $L$ ways of hopping 
to a neighboring
site increasing the number of flipped spins. In addition a   spin flipped in
a   previous hop can be restored. This is immediately seen by applying 
the Hamiltonian to $|v_2\rangle$ in Eq.~(\ref{eq:6}) which can take us back
to $|v_1\rangle$. This latter contribution is cancelled by the term
$-b_n|v_{n-1}\rangle$ in Eq.~(\ref{eq:11}) in the first few steps.  
 Furthermore $a_n/\sqrt{N}\equiv 0$ in the $N\to \infty$ limit, for any finite
$n$. 
We then find that $b_{n+1}=\sqrt{LN}t$ for the first few terms. This is
illustrated in Table \ref{tableI}, where $L=2$. 

If all the elements $b_{n+1}$ would stay constant in the large $N$-limit,
$b_{n+1}=\alpha\sqrt{N}t$ ($\alpha \ge 0$), it is easy to show 
that the lowest state has the
hopping energy $-2\alpha\sqrt{N}t$ and the band gap is
$E_g=U-4\alpha\sqrt{N}t$ for large $U$. In reality, however, after a few
steps the situation becomes more complicated than discussed above.

After three steps, terms proportional to
\begin{equation}\label{eq:16a}
\sum_{mm^{'}m^{''}}\psi^{\dagger}_{2m\uparrow}\psi_{1m\uparrow}
\psi^{\dagger}_{1m^{'}\downarrow}\psi_{2m^{'}\downarrow}\psi^{\dagger}
_{2m^{''}\uparrow}\psi_{1m^{''}\uparrow}|{\rm anti}\rangle,
\end{equation}
enter the calculation.
Since the states with $m<m^{''}$ are equal to the states with
$m>m^{''}$, the norm of the state in Eq.~(\ref{eq:16a}) is
$4N^2(N-1)/2\approx 2N^3$. As a result $b_3$ in Table~\ref{tableI}
is $\sqrt{2.5N}t$ instead of $\sqrt{2N}t$.

After four steps certain states can be reached in two different ways. 
This is illustrated schematically in Fig.~\ref{fig5}. The amplitudes for these 
states then have to be added. The contribution of such a state to $b_{n+1}^2$
is then not $\sim (1^2+1^2)$ as assumed above but $\sim 2^2$. This increases
the value of $b_{n+1}$. In addition, the restoration of a   flipped spin  
can be done in different ways, leading to new states not contained in
$b_n|v_{n-1}\rangle$ in Eq.~(\ref{eq:11}). This is illustrated
in Fig. \ref{fig6}, and it also increases $b_{n+1}^2$.

\noindent
\begin{minipage}{3.375in}
\begin{figure}
  \centerline{\epsfxsize=1.5in \epsffile{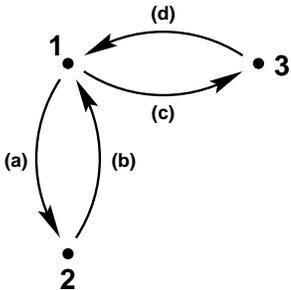}}
  \hspace{2ex}
  \caption[]{\label{fig5}
    The arrows illustrate how the double extra occupancy hops from site 1 
    to 2 and back, followed by a hop to 3 and back. In each hop a spin is 
    reduced. The same final state can be obtained by hops 
    $1\to 3 \to 1\to 2 \to 1$, and the corresponding amplitudes are added.}
\end{figure}
\end{minipage}

\noindent
\begin{minipage}{3.375in}
\begin{figure}
\rotatebox{270}{\centerline{\epsfxsize=0.9in \epsffile{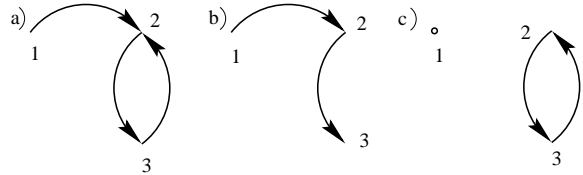}}}
  \hspace{2ex}
  \caption[]{\label{fig6} 
    a) Shows one contribution to $|v_n\rangle$.
    b) and c) show contributions to $H|v_n\rangle$.
    b) Shows how a contribution in $|v_{n-1}\rangle$ can be recovered
       by allowing the extra occupancy to hop back from site 2 to site 3
       and thereby increasing the spins of sites 2 and 3. 
    c) Shows how the extra occupancy instead can hop to site 1, increasing 
       the spins of site 1 and 2 and creating a new state in 
     $|v_{n+1}\rangle$.}
\end{figure}
\end{minipage}

\subsection{Bipartite and non-bipartite systems}

A bipartite system is a system where it is 
possible to partition the system in two sublattices with hopping    
only between the sublattices.   
Nagaoka's theorem\cite{Nagaoka} states that for a such a system with the 
orbital degeneracy $N=1$ and $U=\infty$, the system is ferromagnetic 
when it has one electron or one hole relative to half-filling.
The electron or the hole then have the same hopping possibilities
as in the one-electron
case and it follows that in the large $U$-limit $E_g=U-W$, with the prefactor
exactly equal to one. For $N>1$ or for many of the cases
with $N=1$ but a non-bipartite lattice, where the Nagaoka's theorem has 
not been proven, the reduction of $E_g$ due to hopping is larger than $W$.
This illustrates that Nagaoka's theorem is not valid in these cases. 
Cases with $N=1$ but a non-bipartite lattice are studied below.

For a bipartite lattice where all nonzero  
hopping integrals are equal, we can always construct states which 
are fully bonding or antibonding, by letting the coefficients of the       
orbitals on the two sublattices having the same or the opposite sign. 
This is in general not possible for a non-bipartite lattice.
For instance, for the tetrahedron we found a bonding state with the
energy $3t$ but the antibonding states have only the energy $-t$.
In the Lanczos formalism, this result comes about because an electron can hop
in a closed loop using an odd number of hops. This leads to nonzero 
coefficients $a_n$, which makes the spectrum nonsymmetric around 
$\varepsilon=0$. This also reduces the coefficients $b_{n+1}$ as 
discussed in Sec.
III.D. In the many-body case, the hopping of an electron along a closed
loop usually does not bring us back to the original state, since spins 
have been flipped along the path of the electron. This tends to reduce
the values of $a_n$ and to increase $b_n$, as is illustrated in Table 
\ref{tableI}. Even for $N=1$, hopping may therefore reduce the band gap
{\it more} than one would expect from the one-particle band width. This is 
illustrated in Table \ref{tableV}. In the cases we have studied
the effect is of the order 15-30 $\%$.

\noindent
\begin{minipage}{3.375in}
\begin{table}[h]
  \caption[]{The reduction $U-E_g$ due to hopping for $N=1$ relative to     
    the one-particle band width $W$ for some non-bipartite systems.
    The notations of the systems refer to Fig. \ref{fig2} and ``triang''
     refers to a system with three atoms coupling to each others.}
 \hspace{2ex}
  \begin{tabular}{cc}
   System   &  $\lbrack U-E_g(N=1)\rbrack /W $     \\
\tableline
    c      &  1.25    \\                 
    d     &   1.15    \\                
    e     &   1.19    \\
  triang  &   1.33     \\
  \end{tabular}
  \label{tableV}
\end{table}
\end{minipage}

\section{Interband hopping}
In the previous sections, we have for simplicity assumed that there is
only hopping between orbitals on different sites with the same $m$
quantum number. Here we consider the more general case of hopping
also between different $m$ quantum numbers.
We first consider the step in the Lanczos procedure.
In the one-particle case, this gives  
\begin{equation}\label{eq:i1}
(b_2^{(m)})^2=\sum_{im^{'}}|t_{1m,im^{'}}|^2,
\end{equation}
where the electron is located in orbital $m$ on site 1 in the starting
state. In the many-body case, we start from the state in Eq.~(\ref{eq:5})
and obtain
\begin{equation}\label{eq:i2}
(b_2)^2=\sum_{imm^{'}}|t_{1m,im^{'}}|^2.
\end{equation}
We compare this with the average of the coefficients $(b_2^{(m)})^2$
in the one-particle case 
\begin{equation}\label{eq:i3}
{1\over N}\sum_m (b_2^{(m)})^2={1\over N}(b_2)^2,
\end{equation}
i.e., $b_2$ is typically a factor $\sqrt{N}$ larger  in the many-body    
case than in the one-particle case. This is the same factor as was found
when the hopping between different $m$-quantum numbers was neglected.
The  inter-orbital hopping increases $b_2$ by a factor $\sqrt{N}$ 
in both the one-particle and many-body cases, but leaves their ratio
($=\sqrt{N}$) unchanged.

\noindent
\begin{minipage}{3.375in}
\begin{table}[h]
  \caption[]{The band gap for a six-atom cluster ( d) in Fig. \ref{fig2}).   
     In all cases the absolute value of the hopping matrix elements is $t$. 
     In the ``$sp$''-model we have chosen the signs as if the two orbitals 
     had $s$- ($N=1$) and $p$- character ($N=2$). In the ``rand'' model 
     the signs were chosen randomly. For this model $K=4$ and for the
     ``sp''-model $K$ is effectively somewhat smaller.}
 \hspace{2ex}
  \begin{tabular}{cccc}
     Syst  & \multicolumn{2}{c}{$d(N)=\lbrack U-E_g\rbrack /\lbrack 
W(N)\sqrt{N}\rbrack $} & $d(2)/d(1)$ \\
\tableline
     &   $N=1$  & $N=2$  &  \\
\tableline
   ``$sp$''&  1.15 &  1.31 & 1.14   \\
   ``rand''&  1.23 &  1.10 & 0.89   \\
  \end{tabular}
  \label{tableIV}
\end{table}
\end{minipage}

In Eqs.~(\ref{eq:i1},\ref{eq:i2}) the sign of $t_{im,jm^{'}}$ does 
not matter. When we continue to higher coefficients $b_n$ the sign
becomes important and we have to specify the model in more detail.
We can, for instance, specify the orbitals involved and the geometrical
structure of the lattice. In Table~\ref{tableIV} we show results 
for the lattice d) in Fig.~\ref{fig2}, including a $s$-orbital and 
a $p$-orbital with the lobe along the direction of the three atoms.
This defines the signs of the hopping integrals.     The magnitudes
are assumed to be identical. Alternatively, we can pick the sign
randomly. The latter procedure is appropriate when we want to vary 
$N$, since it is not clear how to define an appropriate set of orbitals
for an arbitrary value of $N$. Table~\ref{tableIV} shows the ratio
\begin{equation}\label{eq:i4}
d(N)={U-E_g(N)\over W(N)\sqrt{N}},
\end{equation}
where $W(N)$ is the one-particle band width for the orbital degeneracy $N$.
In contrast to the intraband model, the one-particle band $W(N)$ width now
depends on $N$.
We also show the ratio $d(2)/d(1)$, which would be unity if the reduction
of band gap due to hopping were proportional to $\sqrt{N}W(N)$. 
This is roughly the case in Table~\ref{tableIV}, but the results are too
limited to allow more definite conclusions.

For the diatomic molecule, discussed in Sec. III.A, and the ``Bethe'' lattice,
discussed in Sec. III.B, we can go to larger values of $N$. The results
are shown in Fig. \ref{fig7}.
These results support a reduction of the band gap due to hopping
which is at least proportional to $\sqrt{N}W(N)$. Although the 
results are still too limited to draw definite conclusions, they
do not support a behavior proportional to $NW(N)$. The $NW(N)$ behaviour 
found for the diatomic molecule with only intraband           
hopping, is therefore probably an artifact of the simple model
used in that case.
 
\noindent
\begin{minipage}{3.375in}
\begin{figure}
  \centerline{\epsfxsize=3in \epsffile{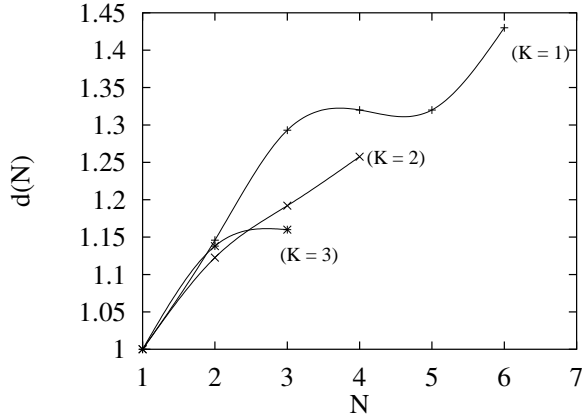}}
  \hspace{2ex}
  \caption[]{\label{fig7} 
   The ratio $d(N)$ (Eq.~(\ref{eq:i4}) between the reduction of the band
gap due to hopping and $\sqrt{N}W(N)$. The figure shows results for 
the diatomic molecule ($K=1$) and the ``Bethe'' lattice with $K=2$ and
$K=3$.}
\end{figure}
\end{minipage}

\section{Discussion}

Above we have studied two types of models, namely models with only intraband 
hopping (between orbitals with the same quantum number $m$)   
and models also with interband hopping.  We have calculated
the reduction of the band gap due to hopping in the large $U$-limit 
for different orbital degeneracies $N$ and for lattices with different 
number $K$ of nearest neighbors.
In the intraband models we find that the hopping contribution grows at least 
as $\sqrt{N}W$, where $W$ is the one-particle band width.
 For $N>K$ there is an interesting crossover, and
the hopping contribution grows as $NW$. This limit should not, however, 
apply to most systems of interest. In the interband models the hopping
contribution seems to be closer to a $\sqrt{N}W(N)$ behavior than a $NW(N)$
behavior even when $K$ is small. Our present results therefore 
support a degeneracy dependence of roughly $\sqrt{N}W(N)$ 
or a somewhat stronger $N$ dependence for most systems. 

It has been suggested that in a strongly 
correlated system the hopping should be strongly suppressed
and it should not reduce the band gap much.   For instance, in models of 
the High $T_c$ compounds, e.g, the $t$-$J$ model, it is found that the 
string of flipped spins in the trace of a moving particle tends to strongly
reduce the dispersion of this particle and that in addition the weight
of the quasi-particle is strongly reduced. This might suggest that hopping
is very inefficient in this case. In the present problem we are, 
however, not interested in the quasi-particle dispersion but in total
energies (Eq.~(\ref{eq:3})). The total energy can be obtained by integrating
the Green's function over frequencies and momenta.\cite{Fetter}
Since the weight of the quasi-particle is strongly reduced in 
correlated systems, this means that the spectral function must have 
substantial weight also for other frequencies. To argue in terms of the
quasi-particle alone neglects important contributions to the total
energy. It is then more convenient to follow the Lanczos approach used here,
which gives an important hopping contribution to the band gap.

It is, nevertheless, true that hopping is suppressed in for large $U$, as
one would expect for strongly correlated systems. The band gap is, however,
an energy difference between ground-state energies, where hopping
has been suppressed differently. Thus hopping is suppressed very strongly
in the $NM$-particle state and less strongly in the $(NM\pm 1)$-particle 
states, so that in the difference (Eq.~(\ref{eq:3})) it appears as if 
the hopping had been enhanced.

The reduced spins along the path of a moving extra occupancy plays two
different roles. One role is that when the extra occupancy moves 
along different paths to a given site, the resulting states depend on 
this path. This is in contrast to the one-electron problem, and it has 
appreciable effects on the Lanczos matrix elements $a_n$ and $b_{n+1}$. 
This is seen in the greater tendency to put weights into the diagonal
elements $a_n$ in the one-particle case, and a reduction of the
one-particle band width for non-bipartite lattices.
For the non-bipartite clusters considered here, the hopping
reduction of the band gap is therefore {\it larger} than the single-particle
band width, even for $N=1$, contrary to the intuitive feeling that
the flipped spins should reduce the hopping contribution (see Table 
\ref{tableV}). 
The second effect is that the
reduction of spins costs energy. This does not enter for $U\to \infty$,
since the energy cost ($\sim t^2/U$) for flipping the spin of one electron  
then goes to zero.
Even for finite values of $U$ this does not enter in the first element
$a_1$, but gradually becomes more important for the higher coefficients.
Normally, however, the spin flip energy is small compared with the
band width $W$. Since the nondiagonal elements $b_{n+1}$ are of the order 
$\sqrt{N}W$ or larger, these elements should tend to dominate the spin-flip
energy, in particular in the large degeneracy limit.

\end{multicols}
\end{document}